# Swar: The Voice Operated PC

**Kamlesh Sharma\*, Maitrei Kohli\*\*, Dr. T. V. Prasad\*\*\***
\*Lingaya`s Institute of Mgmt. & Tech, M.Tech II year student, Faridabad, India,
\*\* Lingaya`s Institute of Mgmt. & Tech., Lecturer, Faridabad, India,
\*\*\* Lingaya`s Institute of Mgmt. & Tech., HOD(CSE), Faridabad, India,

*Abstract -* **Keyboard, although a popular medium, is not very convenient as it requires a certain amount of skill for effective usage. A mouse on the other hand requires a good hand-eye co-ordination. Also current computer interfaces also assume a certain level of literacy from the user. It also expect the user to have certain level of proficiency in English. In our country where the literacy level is as low as 50% in some states, if information technology has to reach the grass root level; these constraints have to be eliminated. As a solution for these, Speech Recognition and hence the concept of Voice operated computer system comes into picture. In this paper we propose a technique to develop a voice recognition system which will be used for controlling computer via speech input from any user i.e. without the use of mouse and / or keyboard. Once developed this system would be of great benefit to physically handicapped people as Instead of scrolling through written procedures on a laptop or handheld computer, they can wear a headset and have their hands and eyes free.**

**Keywords-** data input, voice recognition, Voice Operated, Speech Recognition, HMM.

## I. INTRODUCTION

Keyboard, although a popular medium, is not very convenient as it requires a certain amount of skill for effective usage. A mouse on the other hand requires a good hand-eye co-ordination. It is also cumbersome for entering non-trivial amount of text data and hence requires use of an additional media such as keyboard. Physically challenged people find computers difficult to use. Partially blind people find reading from a monitor difficult. Current computer interfaces also assume a certain level of literacy from the user. It also expected the user to have certain level of proficiency in English. In India the literacy level is as low as 50% in some states. If information technology has to reach the grass root level, these constraints have to be eliminated. Speech interface can help us tackle these problems. [1]

Ever since the invention of the computer scientists and engineers has dreamt of speech recognition and other technologies that would enable human and computer to communicate with each other through spoken language.

Since the early stages of computer developer, engineers and scientists have met with many difficulties in achieving this goal. A lots of research has is going on in this field from long time, however since 1990's, their performance has greatly improved and because of advances in microprocessor and digital signal processing (DSP) they no longer require special hardware. [2]

In this paper we propose a technique to develop a voice recognition system which will be used for controlling computer via speech input from any user i.e. without the use of mouse and / or keyboard. Once developed this system would be of great benefit to physically handicapped people as Instead of scrolling through written procedures on a laptop or handheld computer, they can wear a headset and have their hands and eyes free.

In the next section we proceed to discuss the speech recognition along with its various applications in the same field. The subsequent section then describes the method we intend to use for design the problem. Specific configurations, i.e., the data definitions and the proposed implementation technique have also been discussed.

## II. SPEECH RECOGNITION

Speech recognition refers to the ability to listen (input in audio format) spoken words and identify various sounds present in it, and recognize them as words of some known language. Speech recognition in computer domain may then be defined as the ability of computer systems to accept spoken words in audio format - such as wav or raw - and then generate its content in text format. It involves various steps with issues attached corresponding issues.

The main techniques of implementing speech recognition are:

(i) Voice recording
(ii) word boundary detection
(iii) feature extraction, and
(iv) recognition with the help of knowledge models.

Speech recognition applications include voice dialing (e.g., "Call home"), call routing (e.g., "I would like to make a collect call"), domotic appliance control and content-based spoken audio search (e.g., find a podcast where particular words were spoken), simple data entry (e.g., entering a credit card number), preparation of structured documents (e.g., a radiology report), speech-to-text processing (e.g., word processors or emails), and in aircraft cockpits (usually termed Direct Voice Input).

Various application bibliographies exist, such as Healthcare, military, telephony and other domains, but very few work for people with disabilities. In this proposed scheme we try to device a system that is a style of Human-Machine Interaction in which the user makes voice commands to issue instructions to the machine, and it is thus free from aforementioned difficulties (discussed in section 1).

### III. SYSTEM DESIGN

The prepared system if visualized as a block diagram will have the following components:

Sound Recording and word detection component, feature extraction component, speech recognition component, acoustic and language model.

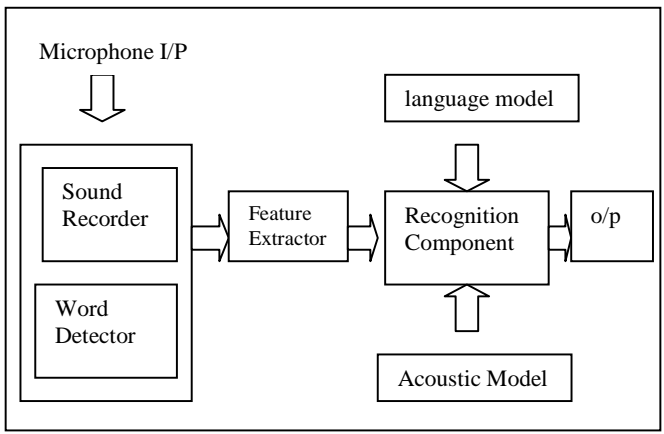

Fig 1: Block diagram of Recognition System

*Sound Recording and Word detection component:*

The component is responsible for taking input from microphone and identifying the presence of words. Word detection is done using energy and zero crossing rate of the signal. The output of this component can be a wave file or a direct feed for the feature extractor. [5]

*Feature Extraction component:*

The component generated feature vectors for the sound signals given to it. It generates Mel Frequency Cepstrum Coefficients and Normalized energy as the features that should be used to uniquely identify the given sound signal.

*Recognition component*

This is a Continuous, Multi-dimensional Hidden Markov Model based component. It is the most important component of the system and is responsible for finding the best match in the knowledge base, for the incoming feature vectors.

*Knowledge Model*

The components consists of Word based Acoustic. Acoustic Model has a representation of how a word sounds. Recognition system makes use of this model while recognizing the sound signal.

Once the training is done, the basic flow can be summarized as the sound input is taken from the sound recorder and is feed to the feature extraction module. The feature extraction module generates feature vectors out of it which are then forwarded to the recognition component. The recognition component with the help of the knowledge model and comes up with the result.

During the training the above flow differs after generation of feature vector. Here the system takes the output of the feature extraction module and feeds it to the recognition system for modifying the knowledge base [1].

### IV. THE ALGORITHM

*Voice Recorder*

- Reads the analog voice signal supplied at the microphone port of the computer, converted to a basic binary code by the ADC (Analog to Digital Converter) built into the computer system.

- Changes this binary input kept in the input buffer to a file format like .au or .wav.

- Prompts the user whether or not he wants to store the buffered data to a sound file with .au or .wav file. If the user says yes, the sound file is created.

*Sound Files Database*

- All the pre – recorded voice samples are assembled here in form of .au or .wav files only. For any kind of an interactions with these files, proper file handling techniques are followed like data buffering, securities and data channeling.

*Comparator*

- Fetches the temporary buffer data gained from the recorder output.

- Compares this data with the data from the sound files database for a proper match by comparing two buffers, one which has data from recorder and other which is connected with the currently open file of sound files database.
- If the match is found, the protected application's access is gained.
- If the match is not found, the user is prompted for the error and the recorder is again initialized.
- This loop continues until the user specifically quits the application in case of match, or aborts the program in case of no match.

## V. IMPLEMENTATION METHODOLOGY

We plan to use a word acoustic model. The system has a model for each word that the system can recognize. The list of words can be considered as language model. While recognizing the system need to know where to locate the model for each word and what word the model corresponds to. This information is stored in a flat file called models in a directory called HMM*s*. Hidden Markov Model (HMM) is a state machine. The states of the model are represented as nodes and the transition are represented as edges. The difference in case of HMM is that the symbol does not uniquely identify a state. The new state is determined by the symbol and the transition probabilities from the current state to a candidate state. Before we can recognize a word we need to train the system. Train command is used to train the system for a new word. The command takes at-least 3 parameters:

• No of states the HMM model should have *N*.
• The size of the feature vector *D*.
• One or more filenames each containing a training set.

For generating an initial HMM we take the *N* equally placed observations (feature vector) from the first training set. Each one is used to train a separate state. After training the states have a *mean* vector which is of size *D*. And a *variance* matrix of size *D x D* containing all zeros. Then for each of the remaining observations, we find the Euclidean distances between it and the mean vector of the states. We assign a observation to the closest state for training. The state assigned to consecutive observations are tracked to find the transitional probabilities.

When a sound is given to the system to recognise, it compares each model with the word and finds out the to model that most closely matches with it. The word corresponding to that HMM model is given as the output.

We propose to implement this system using the .Net API technology. The use of APIs limit the user's prerequisite of DOT NET knowledge required to develop a working project in DOT NET. Simply stating, the users habituated to the use of WINDOWS and DOS platforms can very easily shift to DOT NET by using these. DOT NET supports a number of inbuilt functionalities, which can be very easily embedded into your application. This reduces the size of the code required to make any working application, reducing the time and cost for project development. Also, it has the complete OOP (Object Oriented Programming) paradigm of programming. It supports powerful hardware interactions, necessary for this application. Thus, it is most suited to our needs.

## VI. CONCLUSIONS AND FUTURE WORK

In this paper we present a scheme to develop a voice controlled PC. The key factor in designing such system is the target audience. For example , physically handicapped people should be able to wear a headset and have their hands and eyes free in order to operate the system.

There are several challenges the system needs to deal with in the future. First, the overall robustness of the system must be improved to facilitate implementation in real life applications involving telephone and computer systems. Second, the system must be able to reject irrelevant speech that does not contain valid words or commands. Third, the recognition process must be developed so that commands can be set in continuous speech. And finally, the voice systems must be able to become viable on low-cost processors. Thus, this will enable the technology to be applied in almost any product.